\documentclass{sigchi-ext}
\usepackage[T1]{fontenc}
\usepackage{textcomp}
\usepackage[scaled=.92]{helvet} 
\usepackage{graphicx} 
\usepackage{balance}  
\usepackage{booktabs} 
\usepackage{ccicons}  
\usepackage{ragged2e} 

\usepackage[colorinlistoftodos]{todonotes}

\def\plaintitle{SIGCHI Extended Abstracts Sample File: Note Initial
  Caps} 
\def\emptyauthor{}
\def\plainkeywords{Algorithmic Fairness, Crowdsourcing, Counterfactual}

\title{Measuring Social Biases of Crowd Workers using Counterfactual Queries}

\numberofauthors{4}

\author{%
  \alignauthor{%
    \textbf{Bhavya Ghai}\\
    \affaddr{Stony Brook University} \\
    \affaddr{Stony Brook, USA} \\
    \email{bghai@cs.stonybrook.edu} }  \vfil \alignauthor{%
    \textbf{Q. Vera Liao}\\
    \textbf{Yunfeng Zhang}\\
    \affaddr{IBM Research AI}\\
    \affaddr{Yorktown Heights, USA}\\
    \email{vera.liao@ibm.com} \\
    \email{zhangyun@us.ibm.com} } \vfil \alignauthor{%
    \textbf{Klaus Mueller}\\
    \affaddr{Stony Brook University}\\
    \affaddr{Stony Brook, USA}\\
    \email{mueller@cs.stonybrook.edu}} 
 }
    
\definecolor{linkColor}{RGB}{6,125,233}
\hypersetup{%
  pdftitle={\plaintitle},
  pdfauthor={\emptyauthor},
  pdfkeywords={\plainkeywords},
  bookmarksnumbered,
  pdfstartview={FitH},
  colorlinks,
  citecolor=black,
  filecolor=black,
  linkcolor=black,
  urlcolor=linkColor,
  breaklinks=true,
}

\begin{document}

\copyrightinfo{}

\maketitle
\RaggedRight{} 

\begin{abstract}
Social biases based on gender, race, etc. have been shown to pollute machine learning (ML) pipeline predominantly via biased training datasets. Crowdsourcing, a popular cost-effective measure to gather labeled training datasets, is not immune to the inherent social biases of crowd workers. To ensure such social biases aren't passed onto the curated datasets, it's important to know how biased each crowd worker is. In this work, we propose a new method based on counterfactual fairness to quantify the degree of inherent social bias in each crowd worker. This extra information can be leveraged together with individual worker responses to curate a less biased dataset.    
\end{abstract}



\section{Introduction}
Algorithmic bias occurs when an algorithm exhibits preference or prejudice against certain sections of society based on their identity. It has been termed as the imminent AI danger faced by our society. Its adverse impact has been seen in domains like healthcare, law, education, etc. \cite{o2016weapons}. 
A major source of bias in the ML pipeline arises from the training dataset. Crowdsourcing is widely used to curate training datasets for ML models . Crowdsourced datasets such as MS-COCO, imSitu, etc. have  been found to contain significant social biases \cite{zhao2017men}. One of the major factors contributing to such biases are the labels provided by the crowd workers. If there was a way to identify biased labelers, we could counter their biases by weighing them down or discarding their labels completely.  \\ 

In this work, we propose a novel technique to estimate social biases of each crowd worker. Our approach is based on the idea of counterfactual fairness (CF) which has been previously used to evaluate ML models for fairness. As per CF, a ML model is considered fair if its prediction for an individual and its counterfactual is the same. Here, counterfactual represents the same individual in a hypothetical world with its sensitive attribute i.e. demographic group changed. We have tried to adopt this approach to evaluate crowd workers instead of ML models. Given a labeling task which consists of a sensitive attribute like gender, race, etc., a crowd worker will be considered fair if she provides the same label for a query and its counterfactual. For instance a crowd worker is tasked to label a text statement for toxicity. Assuming we want to gauge gender bias, we will generate the counterfactual query by flipping gender sensitive word(s). A simple example can be "Women are such hypocrites". Its counterfactual will be "Men are such hypocrites". A crowd worker will be considered gender neutral if she provides the same label for the query and its counterfactual.     

The key advantage of this approach is that it integrates seamlessly with the task at hand. Counterfactual queries are added to the existing task in the exact same format as any other query. This leaves the crowd worker unaware that he is being judged for some social bias. So, our hypothesis is that our technique should serve as a better alternative to self reported surveys because it is more resistant to social desirability bias. 
Each crowd worker is evaluated purely based on his own responses to a query and its counterfactual. So, our approach doesn't require any ground truth unlike gold questions technique. This characteristic makes our approach more cost effective to implement for a given new task. It also makes it an elegant solution for subjective tasks like rating a movie where there is no ground truth. Lastly, this property makes our approach immune to the biases of the domain experts who provide labels for gold questions.\\

Crowdsourced labeling tasks can be roughly classified as objective tasks like image classification and subjective tasks like rating a tweet. Quantifying worker bias for objective tasks is relatively easier with the presence of objective ground truth. Different conventional fairness metrics like false positive rate difference, true positive rate difference, etc. can be used by comparing worker's responses against ground truth for gold questions. However for subjective tasks, ground truth is often unavailable or costly to obtain, so such fairness metrics can't be leveraged, even though social biases are more detrimental to these tasks. Hence, in this work we will focus on subjective tasks.
\section{Related Work}
Our work draws its motivation from the domain of Algorithmic Fairness and Crowdsourcing.

\subsection{Crowdsourcing}
Different quality control techniques like attention checks, gold questions, reputation, argumentation etc. are proposed in the literature to identify and control for spammers, noisy annotators, adversarial annotators, etc. In this work, we are interested in identifying and controlling for biased worker. 
We define biased worker as a human annotator who has strong preference or prejudice against a demographic group which are reflected in his/her labels.
One of the most common ways to gauge social biases like gender bias of crowd workers is via self reported surveys . The downside to such surveys is that they suffer from social desirability bias \cite{sdb}. They are usually distinct from the labeling task at hand. Hence, they make the crowd workers conscious that they are being evaluated. 
  Another popular way to measure inherent social bias is Implicit Association test (IAT) . However, recent studies have questioned the effectiveness of this test \cite{iat_counter}.  So, there is a need for a new better alternative to measure social biases.
Furthermore, there seems to be a disconnect between the crowdsourcing and the algorithmic fairness literature. It will be interesting to see conventional fairness metrics being used to measure crowd worker bias and how it compares with the score returned by self reported surveys.

\subsection{Algorithmic Fairness}
The existing literature which focuses on mitigating bias at different stages of the ML pipeline can be broadly classified into 3 stages i.e. pre-processing, in-processing and post-processing . In the pre-processing stage, a given dataset is modified such that the social biases with respect to the sensitive attribute are reduced/removed. In the in-processing stage, novel ML algorithms are devised which return fair predictions even when trained on biased data. Lastly, in the post-processing stage, the predictions of the ML model are modified to make it more fair. \\
Another important stage which needs more attention is the data curation stage. This stage falls before the pre-processing stage. Hence, if the biases are firmly dealt with during data curation, then we won't need to deal with it later on. Multiple factors like skewed representation of a demographic group, biased label distribution from human annotators, etc. can pollute the data curation process. Our work focuses on the data curation stage and aims to improve label quality.   


\section{Our Approach}
\begin{figure}[ht]
    \centering
    \includegraphics[width=0.95\linewidth]{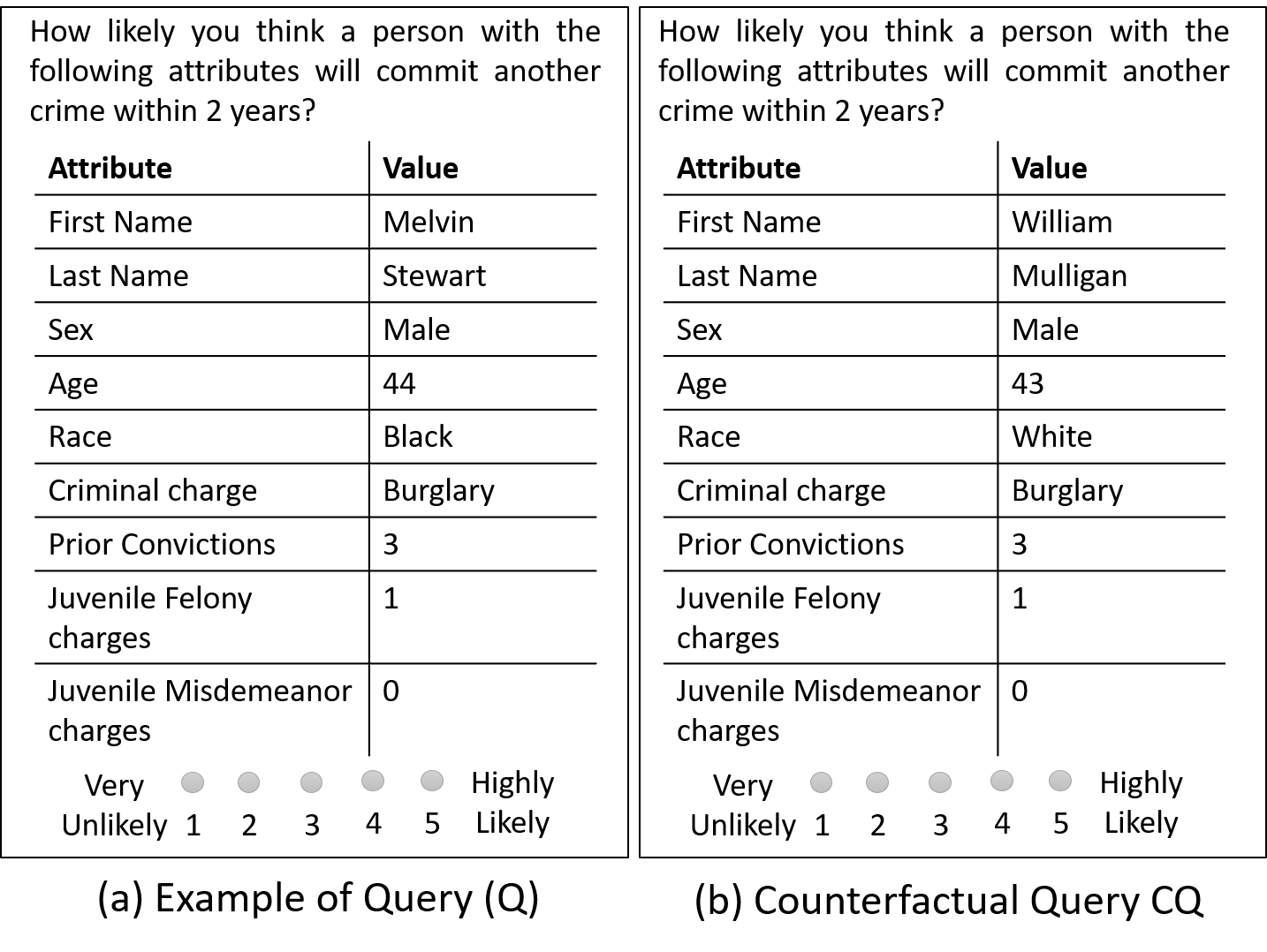}
    \caption{Example of a Query (left) and it's Counterfactual (right) for the recidivism prediction task }~\label{fig:query}
\end{figure}

Let's consider a toy problem where crowd workers are asked to predict recidivism. As shown in fig.\ref{fig:query} (a), a crowd worker is presented with a set of features representing a convict. The task is to predict the likelihood of re-offending within 2 years on a 1-5 scale. Let's say each crowd worker is asked to label x such queries. We will choose a subset of size n out of x queries and generate counterfactuals for only those queries Q.  

In this work, we are considering a narrow class of counterfactuals which can be generated by changing the sensitive attribute and keeping other features constant \cite{garg2019counterfactual}. 
Here, we are dealing with racial bias so we will generate counterfactual query CQ by changing the race attribute (see fig.\ref{fig:query} (b)). 
The mean absolute difference in labels provided for all n pairs of Q and CQ will represent the bias score for a crowd worker as shown eq.\ref{eq:worker_bias}. 
\begin{equation}
    \label{eq:worker_bias}
    Worker Bias = \frac{1}{n} \sum_{i=1}^{n}|Label(Q_i) - Label(CQ_i)|
\end{equation}
A zero score characterizes perfect unbiased behavior and higher values symbolize more biased behavior. This extra information will be used in conjunction with the crowd worker responses to yield fairer labels and hence fairer datasets. A simple way to incorporate this information might be to filter out biased workers whose bias score is beyond a threshold. More sophisticated aggregation algorithms can be adopted as well which better utilize the subtle differences in bias scores. We will evaluate our method by comparing the datasets obtained using our approach and self reported surveys on different fairness metrics.  \\
The key challenge in adopting counterfactual fairness to judge crowd workers relative to ML models is that a crowdworker might relate a previously seen query with its counterfactual. If the crowdworker realizes that she has labeled a very similar query before, she might become conscious that she's being judged. This will defeat our purpose of countering social desirability bias. 
To circumvent this issue, we can play with the ordering of the queries such that a query and its counterfactual are placed far from each other. This will ensure that the memory trace created by the original query gets faded by the time the counterfactual query is encountered.   
Furthermore, we can perturb the counterfactual query by adding small noise to certain numeric features. In the above example, we added noise to the 'Age' feature. Lastly, we can add/modify dummy features like 'First Name', 'Last Name', etc. which are irrelevant to the prediction task. Their role is to carve a slightly different identity for the counterfactual query relative to the original query (see fig. \ref{fig:query}). If the query deals with textual data, we can use different paraphrasing techniques like splitting/combining sentences, substituting synonyms, etc.

\section{Conclusion}
We introduced a new approach to measure social biases of crowd workers using counterfactual queries. 
Next, we plan to test this approach using an empirical study. We want to investigate how the existing techniques fare against different fairness metrics and how does our approach compare against the current state of the art.  
 
 
\bibliographystyle{SIGCHI-Reference-Format}
\bibliography{references}

\end{document}